\documentclass[twoside,twocolumn,9pt]{article}
\usepackage{extsizes}
\usepackage[super,sort&compress,comma]{natbib} 
\usepackage[version=3]{mhchem}
\usepackage[left=1.5cm, right=1.5cm, top=1.785cm, bottom=2.0cm]{geometry}
\usepackage{balance}
\usepackage{mathptmx}
\usepackage{sectsty}
\usepackage{graphicx}
\usepackage{adjustbox} 
\usepackage{lastpage}
\usepackage[format=plain,justification=justified,singlelinecheck=false,font={stretch=1.125,small,sf},labelfont=bf,labelsep=space]{caption}
\usepackage{float}
\usepackage{fancyhdr}
\usepackage{fnpos}
\usepackage[english]{babel}
\addto{\captionsenglish}{%
  \renewcommand{\refname}{Notes and references}
}
\usepackage{array}
\usepackage{droidsans}
\usepackage{charter}
\usepackage[T1]{fontenc}
\usepackage[usenames,dvipsnames]{xcolor}
\usepackage{setspace}
\usepackage[compact]{titlesec}
\usepackage{hyperref}
\usepackage{tablefootnote} 
\usepackage{threeparttable}
\usepackage[normalem]{ulem}

\usepackage{multirow}%
\usepackage{amsmath,amssymb,amsfonts}%
\usepackage{amsthm}%
\usepackage{mathrsfs}%
\usepackage[title]{appendix}%
\usepackage{textcomp}%
\usepackage{manyfoot}%
\usepackage{booktabs}%
\usepackage{algorithm}%
\usepackage{algorithmicx}%
\usepackage{algpseudocode}%
\usepackage{listings}%
\usepackage{siunitx}
\usepackage{epstopdf}

\definecolor{cream}{RGB}{222,217,201}

\begin{document}

\pagestyle{fancy}
\thispagestyle{plain}
\fancypagestyle{plain}{
\renewcommand{\headrulewidth}{0pt}
}

\makeFNbottom
\makeatletter
\renewcommand\LARGE{\@setfontsize\LARGE{15pt}{17}}
\renewcommand\Large{\@setfontsize\Large{12pt}{14}}
\renewcommand\large{\@setfontsize\large{10pt}{12}}
\renewcommand\footnotesize{\@setfontsize\footnotesize{7pt}{10}}
\makeatother

\renewcommand{\thefootnote}{\fnsymbol{footnote}}
\renewcommand\footnoterule{\vspace*{1pt}%
\color{cream}\hrule width 3.5in height 0.4pt \color{black}\vspace*{5pt}} 
\setcounter{secnumdepth}{5}

\makeatletter 
\renewcommand\@biblabel[1]{#1}            
\renewcommand\@makefntext[1]%
{\noindent\makebox[0pt][r]{\@thefnmark\,}#1}
\makeatother 
\renewcommand{\figurename}{\small{Fig.}~}
\sectionfont{\sffamily\Large}
\subsectionfont{\normalsize}
\subsubsectionfont{\bf}
\setstretch{1.125} 
\setlength{\skip\footins}{0.8cm}
\setlength{\footnotesep}{0.25cm}
\setlength{\jot}{10pt}
\titlespacing*{\section}{0pt}{4pt}{4pt}
\titlespacing*{\subsection}{0pt}{15pt}{1pt}

\fancyfoot{}
\fancyfoot[LO,RE]{\vspace{-7.1pt}\includegraphics[height=9pt]{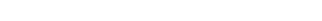}}
\fancyfoot[CO]{\vspace{-7.1pt}\hspace{13.2cm}\includegraphics{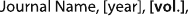}}
\fancyfoot[CE]{\vspace{-7.2pt}\hspace{-14.2cm}\includegraphics{head_foot/RF}}
\fancyfoot[RO]{\footnotesize{\sffamily{1--\pageref{LastPage} ~\textbar  \hspace{2pt}\thepage}}}
\fancyfoot[LE]{\footnotesize{\sffamily{\thepage~\textbar\hspace{3.45cm} 1--\pageref{LastPage}}}}
\fancyhead{}
\renewcommand{\headrulewidth}{0pt} 
\renewcommand{\footrulewidth}{0pt}
\setlength{\arrayrulewidth}{1pt}
\setlength{\columnsep}{6.5mm}
\setlength\bibsep{1pt}

\makeatletter 
\newlength{\figrulesep} 
\setlength{\figrulesep}{0.5\textfloatsep} 

\newcommand{\topfigrule}{\vspace*{-1pt}%
\noindent{\color{cream}\rule[-\figrulesep]{\columnwidth}{1.5pt}} }

\newcommand{\botfigrule}{\vspace*{-2pt}%
\noindent{\color{cream}\rule[\figrulesep]{\columnwidth}{1.5pt}} }

\newcommand{\dblfigrule}{\vspace*{-1pt}%
\noindent{\color{cream}\rule[-\figrulesep]{\textwidth}{1.5pt}} }

\makeatother

\twocolumn[
  \begin{@twocolumnfalse}
{\includegraphics[height=30pt]{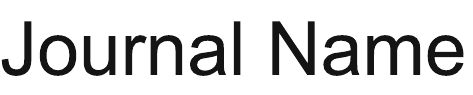}\hfill\raisebox{0pt}[0pt][0pt]{\includegraphics[height=55pt]{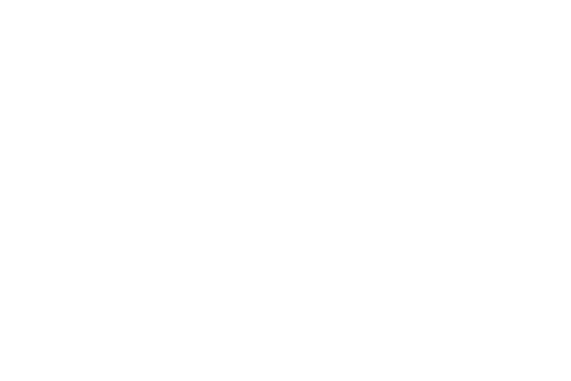}}\\[1ex]
\includegraphics[width=18.5cm]{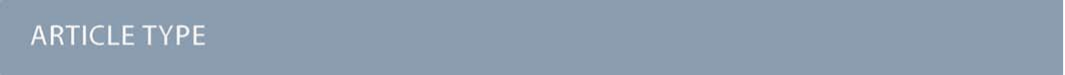}}\par
\vspace{1em}
\sffamily
\begin{tabular}{m{4.5cm} p{13.5cm} }

\includegraphics{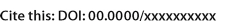} & \noindent\LARGE{\textbf{A termolecular reaction theory for gas-phase nucleation based on long-range intermolecular forces\dag}} \\
\vspace{0.3cm} & \vspace{0.3cm} \\

 & \noindent\large{Yu Wang,\textit{$^{a,\ddag}$} Arnab Choudhury,\textit{$^{b,\ddag}$} Felix Graber,\textit{$^{b,\ddag}$} Ruth Signorell\textit{$^{b}$} and Jes\'us P\'erez-R\'ios\textit{$^{a,*}$}} \\

\includegraphics{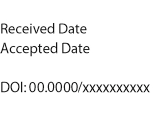} & \noindent\normalsize{The birth of a new phase is usually described thermodynamically, but in the gas phase it begins as chemistry. Here, we show that gas-phase nucleation can be described as a termolecular reaction network controlled primarily by long-range intermolecular forces. In single-component mixtures, dimer formation emerges as a direct termolecular process, whereas in binary mixtures a chaperon mechanism can enhance nucleation, with the second component acting as a catalyst for cluster formation. By incorporating cluster evaporation, the same framework can be extended beyond the collision limit, providing a molecular route to nucleation in regimes where larger critical clusters become relevant. We test the theory against unary and binary nucleation of water, toluene, and butane, obtaining agreement in absolute rates within one order of magnitude across the explored temperature and density ranges. These results identify long-range intermolecular forces as molecular drivers of gas-phase nucleation and establish elementary termolecular chemistry as a predictive route to cluster formation.

} \\

\end{tabular}

 \end{@twocolumnfalse} \vspace{0.6cm}

  ]

\renewcommand*\rmdefault{bch}\normalfont\upshape
\rmfamily
\section*{}
\vspace{-1cm}

\newcommand{\blue}[1]{\textcolor{blue}{#1}}

\footnotetext{\textit{$^{a}$~Department of Physics and Astronomy, Stony Brook University, Stony Brook 11794, NY, USA}}
\footnotetext{\textit{$^{b}$~Department of Chemistry and Applied Biosciences, Laboratory of Physical Chemistry, ETH Z\"urich, Vladimir-Prelog-Weg 2, CH-8093 Z\"urich, Switzerland}}
\footnotetext{\ddag~Y.W., A.C., and F.G. contributed equally to this work.}
\footnotetext{*~Corresponding author. Email: jesus.perezrios@stonybrook.edu}
\footnotetext{\dag~Supplementary Information (SI) available: experimental conditions, additional figures, tables of rate coefficients, and details of the long-range coefficients. See DOI: 00.0000/00000000.}

\section{Introduction}

Nucleation underlies first-order phase transitions, explaining how a new phase emerges from a parent phase. Nucleation is largely treated from a macroscopic perspective, as in the celebrated classical nucleation theory (CNT)~\cite{Kalikmanov2013,BeckerDoring1935,Zeldovich1943,VolmerWeber1926}. In the CNT, nucleation is viewed as an activation process: once a cluster exceeds a given threshold, particles spontaneously accrete, and the cluster grows. The cluster is treated thermodynamically, using bulk properties of both phases, characterized by surface tension and chemical potential. However, as more experimental platforms have been developed to characterize critical clusters in gas-phase nucleation, growing evidence indicates that CNT falls short in predicting nucleation rates, especially its temperature dependence~\cite{Rudek_1999,Strei_1986,Manka_2010,Lihavainen_2001,Luijten_1997,Hruby_1996,Krohn2020,Mullick_2015,Wolk_2001,Brus_2008,Gharibeh_2005,Iland_2004}. Classical nucleation theory can be improved by incorporating the curvature dependence of the surface tension, yielding better results~\cite{cCNT}, although this theory requires detailed knowledge of the equation of state and is only valid for clusters containing hundreds of particles or more~\cite{limits_cCNT}.

On the other hand, molecular dynamics simulations are the key microscopic tool to characterize the onset of nucleation~\cite{YasuokaMatsumoto1998,TanakaEtAl2005,DiemandEtAl2013}, yielding quantitative results. Similarly, it is possible to study nucleation phenomena using master-equation approaches---fully quantum mechanical methods~\cite{Klippenstein}. Although both approaches are microscopic and can elucidate the nucleation dynamics of specific systems, their complexity limits mechanistic insight and transferability across systems.

Here, we introduce a first-principles molecular theory of gas-phase nucleation based on a network of elementary termolecular reactions. The central idea is that the formation of the first stable clusters can be treated as a direct three-body recombination process, using a reaction-dynamics framework rooted in few-body physics. In this approach, the rate constants are controlled primarily by readily accessible long-range intermolecular interactions. The framework predicts that, in single-component mixtures, nucleation is dominated by direct termolecular formation of dimers, whereas in two-component mixtures an additional chaperon channel can enhance the nucleation rate, with the second species acting as a catalyst for cluster formation. We test the theory against experimentally retrieved nucleation rates for gas mixtures containing butane, toluene, water, and CO$_2$~\cite{Chakrabarty-2017,Li2021,Krohn2020,Feusi2023,Choudhury2026} at very high supersaturation, where nucleation is dominated by dimer formation, i.e., in the collision limit. The model predicts absolute rates within one order of magnitude of the experimental results, demonstrating its accuracy and predictive power. By incorporating Arrhenius-like rate constants for cluster evaporation, the same reaction-network framework can be extended beyond the collision limit, as confirmed by temperature- and pressure-dependent nucleation rates for CO$_2$~\cite{Dingilian2021}. Thus, the present theory provides a transferable molecular-level framework for predicting gas-phase nucleation from elementary reaction dynamics and long-range intermolecular forces.


\section{Termolecular reaction network for gas-phase nucleation}

In the collision limit (barrier-free nucleation), nucleation kinetics is determined by the kinetics of dimer formation, which under these conditions is the critical cluster. Molecular-level experimental nucleation data have recently become available through an approach that combines Laval expansions, which initiate cluster formation under uniform conditions, with mass spectrometric detection~\cite{Ruth_1,Li2021}. The gas mixtures considered here contain a carrier gas M and either one condensable vapor A or two vapor species A and B, referred to as single-component (unary) and two-component (binary) mixtures, respectively.

\begin{figure*}[]
    \centering
    \includegraphics[width=1\linewidth]{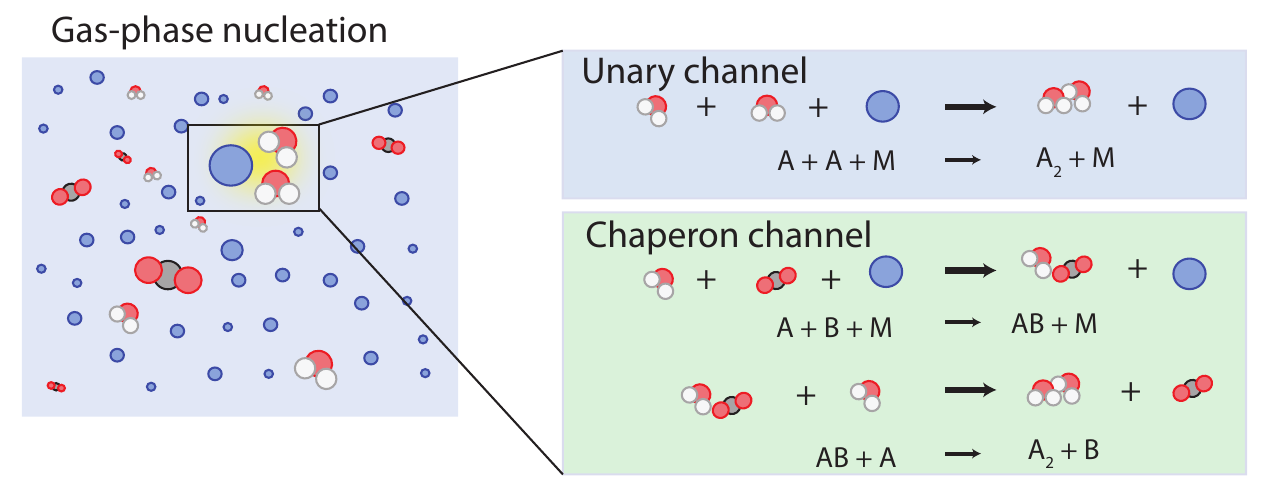}
    \caption{Gas-phase nucleation as an elementary termolecular reaction network. Schematic representation of gas-phase nucleation in a dilute mixture containing a condensable vapor A, a carrier M, and, in binary mixtures, a second vapor species B. The highlighted region illustrates the local molecular event that initiates nucleation: the formation of the first stable cluster through a three-body encounter. The zoomed-in region shows the elementary termolecular reaction pathways underlying the nucleation mechanism. In single-component mixtures, nucleation proceeds through the unary channel, A + A + X $\rightarrow$ A$_2$ + X, in which the third body removes excess energy and stabilizes the dimer. In binary mixtures, an additional chaperon channel can operate, in which the second component B first forms an intermediate cluster AB, A + B + Y $\rightarrow$ AB + Y, that subsequently reacts with A to yield the dimer, AB + A $\rightarrow$ A$_2$ + B. }
    \label{fig1}
\end{figure*}

We assume that nucleation in the collision limit results from a network of termolecular reactions, as shown in Fig.~\ref{fig1}. All termolecular reactions considered are three-body recombination in which three reactants collide almost simultaneously, creating a bound state between two of them as the reaction product, while the third reactant carries away the excess energy. In a single-component mixture, or unary nucleation, condensation of gas A proceeds through dimer formation (A$_2$) as the rate-determining step. The main reaction to consider is the unary channel:
\begin{equation}
\label{r1}
\ce{A + A + X ->[k^u_{3,X}] A_2 +X},
\end{equation}
where X can be A, B (in two-component mixtures only), or M, acting as the molecule that carries away the excess energy. Since M is present in excess, it will be the dominant contribution to the unary channel. The contribution of reaction~\ref{r1} to the nucleation rate is
\begin{equation}
\label{unary}
J_{\text{unary}}=\left(k^u_{3,A}[A][A]+k^u_{3,B}[A][B]+k^u_{3,M}[A][M]\right)[A],
\end{equation}
with the brackets indicating concentrations. 

In general, the interaction between the two relevant vapors A and B is stronger than their interactions with the carrier gas M, which is comparatively weakly interacting. As a result, a new reaction pathway opens, as shown in Fig.~\ref{fig1}. It consists of a termolecular reaction
\begin{equation}
\label{r2}
\ce{A + B + Y ->[k^b_{3,Y}] AB + Y},
\end{equation} 
where Y can be M or B, followed by a reaction of the weakly bound AB cluster with the nucleating vapor A via a bimolecular reaction
\begin{equation}
\label{r3}
\ce{AB + A ->[k^b_2] A_2 +B},
\end{equation}
again forming an A$_2$ dimer. We refer to this channel as the chaperon channel. We assume that the rate-determining step is given by reaction~\ref{r2}, which implies that reaction~\ref{r3} is highly efficient~\cite{Li2021}. B molecules can be considered catalysts of nucleation, since they actively participate in the nucleation mechanism but are not consumed. The contribution of the chaperon channel to the nucleation rate is given by 
\begin{equation}
\label{Chaperon}
J_{\text{Chaperon}}=\left(k^b_{3,M}[B][M]+k^b_{3,B}[B][B]\right)[A].
\end{equation}
Again, as M is present in excess, it will usually be the dominant contribution. The total nucleation rate is given by the unary plus the chaperon contribution:
\begin{equation}
J=J_{\text{unary}}+J_{\text{Chaperon}}.
\end{equation}

\section{Few-body reaction dynamics for gas-phase nucleation}

The reaction rates for reactions (\ref{r1}) and (\ref{r2}) are calculated within a classical capture model, following the methodology introduced in Refs.~\cite{mirahmadi2021classical,Yu2023}. Briefly, our capture model treats termolecular reactions within a direct approach, without invoking intermediate complexes, by assuming that the reaction occurs during the simultaneous collision of the three reactants. This approach is justified in light of previous results for several systems, including ozone formation, sulfur recombination, halogen recombination, and ion-atom recombination reactions~\cite{Sulfur,Ozone,perez2014comparison,JPR_2018,Halogens,Koots2026}. In addition, for the reactions under consideration, the interactions in the intermediate complexes are mostly van der Waals in character, leading to a low density of states and therefore short lifetimes. The main idea of any capture model is to define a capture radius such that, once the reactants cross it, the reaction occurs with 100\% probability. In our capture model, we assume that long-range interactions between the reactants determine the reaction course.

The interaction energy landscape of the termolecular reaction X + Y + Z can be mapped onto a six-dimensional potential energy surface, which we describe using hyperspherical coordinates. As is typical in atom-recombination reactions, we assume that the hyperradius, equivalent to the radius of a higher-dimensional sphere, is the reaction coordinate. This mapping enables the application of a capture model in which reaction occurs when the collision energy is sufficient to overcome the effective centrifugal barrier. For a given collision energy, this condition defines a maximum impact parameter, which is then used to obtain the termolecular reaction rate~\cite{mirahmadi2021classical}:
\begin{equation}
\label{eq_rate2}
k_{3}(T)=\frac{4\pi^3}{3\Gamma(1/3)\sqrt{\mu_3}}(2C_{6}^{\mathrm{eff}})^{5/6}(k_BT)^{-1/3},
\end{equation}
where $k_B$ is the Boltzmann constant, $T$ is the temperature, and $C_{6}^{\mathrm{eff}}$ is the effective long-range interaction coefficient, as explained in the SI. In this equation, $\Gamma(x)$ represents the Gamma function of argument $x$, and $\mu_3$ is the three-body reduced mass, given by
\begin{equation}
\mu_3=\sqrt{\frac{m_\text{X} m_\text{Y} m_\text{Z}}{m_\text{X} + m_\text{Y} + m_\text{Z}}}.
\end{equation}
A detailed derivation can be found in the SI.
\section{Experiment}

A more detailed description of the experimental setup is provided in the SI (Fig. S3) and Refs.~\cite{Ruth_1, Li_2019}. Briefly, gas mixtures of nucleating compounds (butane, toluene, water, and CO$_2$) immersed in a carrier gas (consisting of Ar, N$_2$ or mixtures thereof) were supplied to the stagnation volume of a Laval nozzle. The expansion of the gas mixture through the Laval nozzle is accompanied by a very rapid decrease in temperature and pressure, which leads to supersaturation of the nucleation compound and thus to nucleation in the post-nozzle flow of the expansion (see Fig. S3). Temperature and pressure in the post-nozzle flow -- and thus supersaturation -- are constant (uniform conditions). The supersaturations of the nucleating species were calculated using the Wagner Equation and coefficients from Ref.~\cite{VDIWaermeatlas2019} for equilibrium vapor pressures. The first unique feature of our experiment is the initiation of nucleation under uniform conditions, which is a prerequisite for well-controlled nucleation experiments as nucleation is exquisitely sensitive to changes in conditions. The distinctive capability is the ability to observe critical clusters at a truly molecular level, yielding information on the entire cluster distribution, individual cluster sizes (number of molecules per cluster), their chemical composition and their absolute number concentrations. All these factors are essential for determining critical clusters, nucleation rates, and information about potential mechanisms directly from experimental data (see e.g. Refs.~\cite{Li_2019, Nucleation_review2}). This is achieved by probing the post-nozzle flow with a home-built mass spectrometer after soft ionization of the clusters with a home-built vacuum ultraviolet (VUV) laser with a photon energy (here 13.8 eV) just above the lowest ionization energy. This detection method offers the necessary sensitivity and leaves the clusters essentially intact. The specific conditions and nucleation rates for the different experiments reported here are listed in the SI in Tables S3 to S10. The conditions were chosen so that the concentration of the carrier gas (M) exceeds that of the condensable gas (A) by several hundred to several thousand times. Furthermore, the catalyst B (Eq. \ref{r2}, here B = CO$_2$) was also present in excess compared to the condensable gas. Typically, it exceeded the concentration of A by up to a few ten times. It is also important to note that the catalyst does not condense under these conditions.

\section{Results and discussion}

\subsection{Single-component mixtures and temperature effects}

\begin{figure}[h]
    \centering
    \includegraphics[width=1\linewidth]{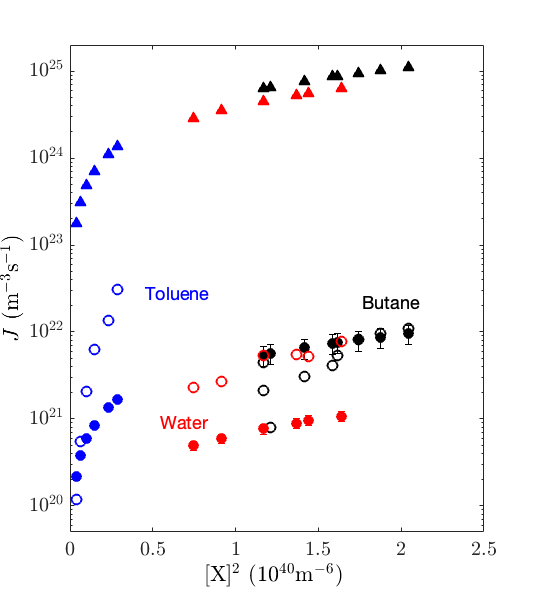}
    \caption{Unary nucleation rates for butane, toluene, and water as a function of the square concentration of the nucleating vapor for 51, 55, and 57~K, respectively. The hollow circles represent the experimental data, the solid circles show our theoretical results, and the triangles show the results from the hard-sphere model. The blue symbols refer to toluene, the black to butane, and the red to water. The error bars for the theory calculations account for the uncertainty in determining the effective long-range coefficient, as explained in the SI. Note that for water and toluene, the theoretical error bars are smaller than the size of the symbols. The experimental data for water and toluene are newly measured. The butane data were taken from Ref.~\cite{Choudhury2026}.}
    \label{fig2}
\end{figure}

The results for nucleation rates in single-component mixtures of butane (black symbols), toluene (blue symbols), and water (red symbols) as a function of the squared concentration of the respective nucleating species (X = butane, toluene, and water, respectively) are shown in Fig.~\ref{fig2}. The empty circles represent experimental rates, and the filled circles are the results obtained with our model assuming the same conditions (partial pressures and temperature) as in the experiment (see Tables S3--S5 in the SI). The theoretical unary nucleation rates were calculated from Eq.~\ref{unary} with rates obtained from Eq.~\ref{eq_rate2} using the effective long-range coefficients given in the SI. Theoretical predictions and experimental results are in excellent agreement. The maximum deviation (for toluene) is within a factor of 10, which still lies within the estimated experimental uncertainties for the absolute value of the nucleation rate~\cite{Krohn2020}. (Note that within a measurement series, we estimate uncertainties between two data points of about a factor of two.) The theoretical calculations include an error bar due to approximations used to determine the pairwise van der Waals coefficient (see SI). 

A more detailed examination of the results shows that the agreement is best for butane (for all data points typically within less than a factor of approximately 2). The theoretical predictions for water data are systematically higher than the experimental data (by approximately a factor of 4). For toluene, the deviation increases with increasing toluene monomer concentration from almost zero to about a factor of 10. As already mentioned, these agreements are excellent considering the given experimental and theoretical uncertainties. It should be noted that other, more established nucleation theories, such as classical nucleation theory~\cite{Rudek_1999,Strei_1986,Manka_2010,Lihavainen_2001,Luijten_1997,Hruby_1996,Krohn2020,Mullick_2015,CNT_vs_exp} and mean-field kinetic nucleation theory~\cite{Krohn2020,Lippe_2018,Bennett_2012}, can only predict nucleation rates at best within two to three orders of magnitude (in some cases even much worse) of the experimental data.

Hard sphere models are often used to calculate capture cross sections. To demonstrate the differences in the results of our termolecular capture model, we implemented a simple hard sphere model (see SI Section S2.1). The nucleation rates obtained from the hard sphere model are indicated in Fig.~\ref{fig2} by the filled triangles. These systematically overestimate the nucleation rates by several orders of magnitude, clearly showing that, unlike our model, they lack predictive power. This behavior has been previously reported and is a pathological issue of the hard-sphere model~\cite{Krohn2020,Lippe_2018,Feusi2023,Li_2019}.

\begin{figure}
    \centering
    \includegraphics[width=1\linewidth]{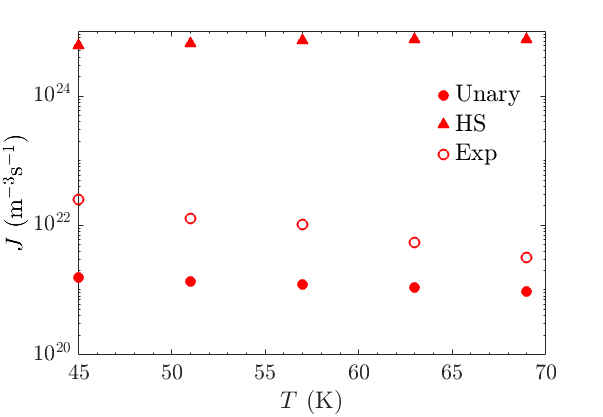}
    \caption{Temperature-dependent water nucleation rate for a given fixed water density. The experimental data were obtained by keeping the water density as close as possible to $1.33\times 10^{20}$~m$^{-3}$ and have not been published before. The unary and hard-sphere (HS) results are calculated for the specific water density. }
    \label{fig3}
\end{figure}

The nucleation of any substance is controlled by the system's physical conditions, namely, temperature and partial pressures. Above, we investigated the influence of the partial pressure of the condensable substance on the nucleation rate at a given temperature. Here, we briefly address the effects of temperature on nucleation. Fig.~\ref{fig3} shows the unary nucleation rate for water over the temperature range from 45 to 75~K for a given water monomer concentration (equivalent results for two further water concentrations are documented in Fig. S2 in the SI). The experimental data (empty circles) show a consistent negative slope in the nucleation rate with respect to temperature. This behavior is characteristic of barrierless chemical reactions, such as three-body recombination. In very good agreement with the experimental trend, our results using the unary reaction channel (filled circles) predict a negative slope for the nucleation rate with respect to temperature, though slightly less steep. Therefore, the experimental data can be viewed as additional evidence that nucleation in a highly supersaturated environment mainly results from a three-body recombination. The hard sphere model (filled triangles), on the other hand, not only drastically overestimates the rates, but also shows the opposite temperature trend, i.e., an increase in the rate with rising temperature.

\subsection{Two-component mixtures}

\begin{figure*}[t]
    \centering
    \includegraphics[width=1\linewidth]{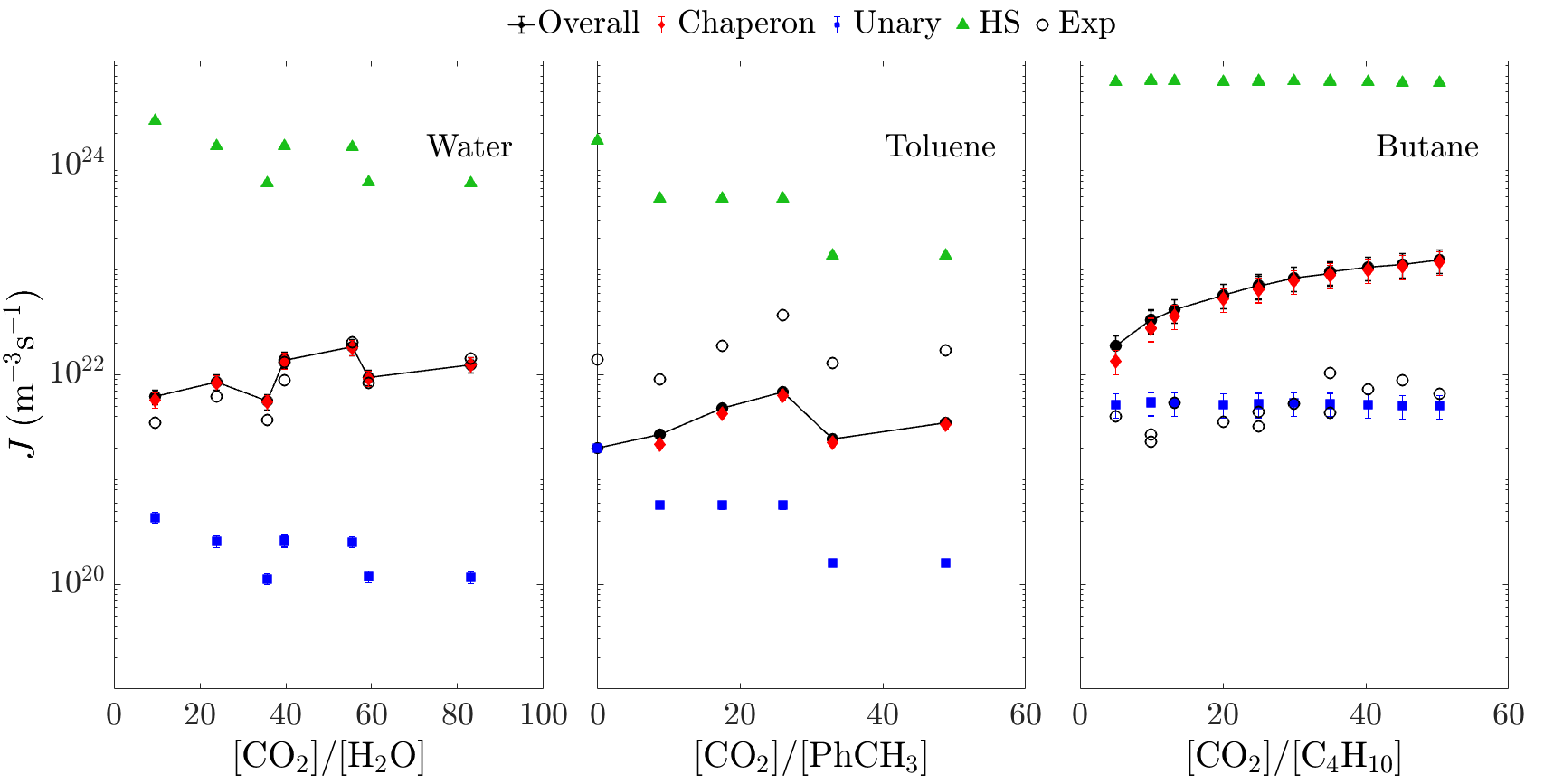}
    \caption{Binary nucleation rates for water, toluene, and butane. The left panel shows the water binary nucleation rate at 57~K as a function of the concentration ratio between CO$_2$ and water (experimental data from Ref.~\cite{Feusi2024}). The middle panel shows the toluene binary nucleation rate at 55~K as a function of the concentration ratio between CO$_2$ and toluene (experimental data from Ref.~\cite{Li2021}). The right panel shows the butane binary nucleation rate at 51~K as a function of the concentration ratio between CO$_2$ and butane (experimental data from Ref.~\cite{Choudhury2026}). Empty black circles represent the experimental results, blue squares show the theoretical results assuming only unary reaction channels, and red diamonds denote the chaperon reaction channel. The total theoretical nucleation rate, given as the sum of the unary and chaperon channels, is shown by filled black circles connected with a line. These points are mostly obscured by the chaperon results because this is the dominant reaction channel. Green triangles represent nucleation rates obtained from the hard-sphere model (HS).}
    \label{fig4}
\end{figure*}

We have previously demonstrated experimentally that the presence of CO$_2$ accelerates the nucleation of water and toluene, and also of butane to a minor extent~\cite{Li2021, Feusi2024, Choudhury2026}. CO$_2$ therefore acts as an efficient nucleation accelerator (or catalyst) for these compounds in two-component gas mixtures. Fig.~\ref{fig4} shows a comparison between the experimental nucleation rate (empty black circles) and the predictions from our termolecular reaction model for water, toluene, and butane as a function of the ratio of the concentrations of CO$_2$ and the nucleating species. The calculated overall reaction rates (i.e., the sum of the unary and chaperon channels) are indicated by filled black circles and the black line, the unary rates are indicated by filled blue squares, and the chaperon rates are indicated by red diamonds. All values are given in the SI in Tables S6-S9. As in the unary cases, we find very good agreement between the predicted overall rates and the experiment, confirming the predictive power of our model. In addition, we show hard-sphere rates (filled green triangles), which as in the case of unary nucleation overestimate the experimental rates by orders of magnitude.

\begin{table*}[]
	\begin{center}
	\caption{Three-body recombination rate coefficients for unary and chaperon reaction channels for water, toluene, and butane nucleation. The reaction rate coefficients are expressed in cm$^6$/s.}\label{tab:rates}
    
\begin{tabular}{c|c|ccc|cc}\toprule
 & & \multicolumn{3}{c}{Unary: A + A + X (Eq. (1))} & \multicolumn{2}{c}{Chaperon: A + CO$_2$ + Y (Eq. (3))}\\
\bottomrule
Nucleating species A & Temperature (K)& X=Ar & X=N$_2$ & X=CO$_2$ & Y=Ar & Y=N$_2$ \\
\bottomrule
 C$_4$H$_{10}$& 51& $7.9\times10^{-30}$ & $9.8\times10^{-30}$ & $9.6\times10^{-30}$ & $4.2\times10^{-30}$ & $4.9\times10^{-30}$   \\
 C$_7$H$_{8}$ &  55 & $1.0\times10^{-29}$ & $1.3\times10^{-29}$ & $1.1\times10^{-29}$ & $4.5\times10^{-30}$& $5.2\times10^{-30}$\\
 H$_2$O  & 57 & $1.2\times10^{-30}$ &  $1.4\times10^{-30}$ &  $1.8\times10^{-30}$ &  $1.8\times10^{-30}$ &  $1.9\times10^{-30}$ \\
 \bottomrule
\end{tabular}
	\end{center}
    
\end{table*}
The comparison between the calculated overall rate and the unary and chaperon rates reveals for all three compounds a predicted overall rate that is dominated by the chaperon contribution (the filled black circles almost completely overlap with the red diamonds). The unary channel (filled blue squares) is predicted to play a minor role only. Since the calculated rate constants for the different channels (see Table~\ref{tab:rates}) differ by less than a factor of three for all three systems, this is largely due to the fact that CO$_2$ is present in excess compared to the condensable substance A (see abscissa in Fig.~\ref{fig4}). For this reason, the contribution of the unary channel A + A + Ar/N$_2$ is always lower than the contribution of the chaperon channels A + CO$_2$ + Ar/N$_2$. Furthermore, the contribution of the unary channel A + A + CO$_2$ is negligible compared to the unary channel A + A + Ar/N$_2$, as the concentration of CO$_2$ is about an order of magnitude lower than the concentration of Ar/N$_2$. The chaperon channel plays a larger role for water than for toluene or butane -- an effect that is also influenced by the rate constants (Table~\ref{tab:rates}). For water, the rate constants for A + CO$_2$ + Ar/N$_2$ and A + A + Ar/N$_2$ are almost identical, while in the case of toluene and butane the latter exceed the former by a factor of about 2. This behavior correlates with water being a small molecule with a dipole moment of 1.68~Debye, so that the dipole-induced dipole interaction controls the water-CO$_2$ long-range interaction. In the case of butane and toluene, the van der Waals coefficient is dominated by the induced-dipole-induced-dipole interaction due to the large size of the molecules. For that reason, our calculation shows that CO$_2$ catalyzes water nucleation more efficiently, followed by toluene nucleation (dipole of 0.38~D) and butane nucleation (no dipole moment).  

\begin{figure}
    \centering
    \includegraphics[width=1\linewidth]{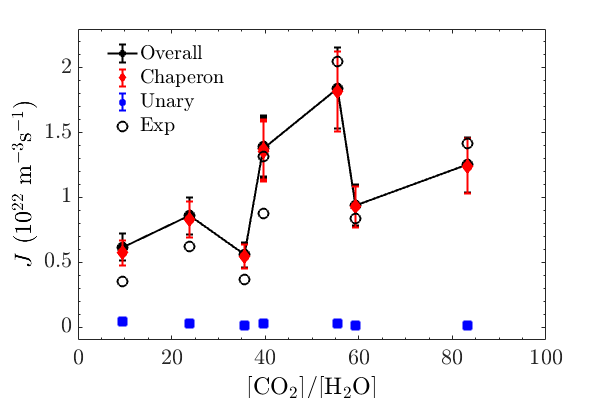}
    \caption{ Water binary nucleation rate at 55~K as a function of the concentration ratio between CO$_2$ and water. The empty black circles represent the experimental results (data from Ref.~\cite{Feusi2024}); the blue squares show the results from our theory assuming only unary reaction channels; and the red diamonds denote the chaperon reaction channel, which overlaps with the experimental results. The total theoretical nucleation rate, given as the sum of the unary and chaperon channels, is shown by filled black circles connected with a line. These points are mostly obscured by the chaperon results because this is the dominant reaction channel.}
    \label{fig5}
\end{figure}

For water and toluene, the calculated overall rate agrees remarkably well with the experimental rates (black empty circles). In particular, the theory reproduces the behavior of the reaction rate as a function of the ratio of CO$_2$ to the concentration of substance A with high fidelity. This is again visualized for the water system in Fig.~\ref{fig5} using a linear scale for the rates. The theory shows that the dependence of the rates on the concentration ratio is dominated by the chaperon channel, providing strong evidence that the molecular model captures the behavior of water and toluene nucleation. For butane, the agreement between experiment and theory is less quantitative. In the experiment, we observed only a minor contribution to the rate from the chaperon channel \cite{Choudhury2026}. Therefore, the fact that the experimental butane data agree better with the predicted unary rate (filled blue squares) than with the overall rate implies that the model overestimates the contribution of the chaperon channel. The model assumes that the reaction in Eq.(\ref{r2}) forms a chaperon complex AB that is long-lived enough to form A$_2$ dimers (Eq.~(\ref{r3})) with a probability of 100\%. This assumption appears to overestimate the effectiveness of the chaperon channel in the case of butane. This could be related to the fact that butane has no dipole moment, indicating that, for polar vapors, the chaperon channel is especially relevant, whereas for apolar substances the unary channel may dominate.

\subsection{Beyond the collision limit}

\begin{figure}
    \centering
    \includegraphics[width=1\linewidth]{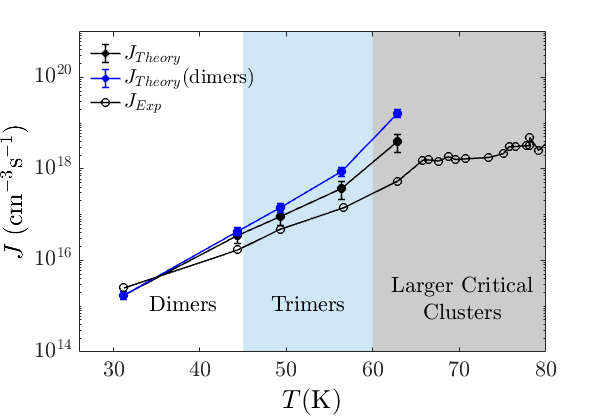}
    \caption{Unary CO$_2$ nucleation rate as a function of the temperature. The experimental data taken from Ref.~\cite{Dingilian2021} have been thermally corrected following Feder et al. \cite{Feder1966}. The filled blue circles represent the results from Eq.~(\ref{unary}), whereas the filled black circles correspond to Eq.~(\ref{eqJ}), which includes dimer evaporation. Each shaded region is labeled with the size of the critical cluster. The nature of the theoretical error bars are explained in the SI (Section S2.2). }
    \label{fig6}
\end{figure}

Depending on the system's physical conditions, the critical cluster size may exceed the dimer. In these scenarios, a termolecular reaction alone does not suffice to explain the onset of nucleation, and evaporation must be included. Our model can accommodate multiple reaction layers within the reaction network, accounting for further termolecular processes leading to larger cluster sizes and evaporation. To illustrate the methodology, we focus on the unary nucleation of CO$_2$, as this is the only system for which experimental data are available at the molecular level, ranging from the collision limit (dimer as critical cluster) to barrier-controlled nucleation (critical clusters larger than the dimer) \cite{Krohn2020, Dingilian2021}. Experimental data up to a temperature of 80 K recorded at CO$_2$ partial pressures spanning three orders of magnitude are shown in Fig.~\ref{fig6} as empty black circles (see also Table S10 in the SI). For temperatures between about 45 and 65 K, the CO$_2$ trimer was predicted to be the leading critical cluster (blue area), while above this value larger critical clusters (gray area) will become important. Here we show, using the trimer region as an example, how our approach can be extended to larger critical clusters. To describe such a scenario, we need to account for the reaction network up to the CO$_2$ trimer:  
\begin{eqnarray}
\label{eq19}
\ce{CO2 + CO2 + M <-->[k_3][k_d] (CO2)2 +M} \nonumber \\
\ce{(CO2)2 + CO2 + M ->[\hat{k}_3] (CO2)3 +M}.
\end{eqnarray}
The first reaction represents a termolecular reaction with reaction coefficient $k_3$ in the forward direction. The backward direction is a bimolecular reaction for dissociation or cluster evaporation with a reaction rate $k_d$. The second reaction is a termolecular reaction with reaction rate coefficient $\hat{k}_3$ for the formation of $(\text{CO}_2)_3$, the critical cluster size. Assuming a steady state for the dimer concentration, we find that the unary nucleation rate is given by
\begin{equation}
\label{eqJ}
J=\frac{\hat{k}_3(T) k_3(T)[\text{CO}_2]^3[M]}{k_d(T)+\hat{k}_3(T)[\text{CO}_2]}.
\end{equation}
The main difference between the nucleation rate from Eq.~(\ref{eqJ}) and that from Eq.~(\ref{unary}) is the presence of a bimolecular dissociation process. Here, we assume an Arrhenius-like rate 
\begin{equation}
k_d(T)= 4\pi R_{\text{CO}_2}^2\sqrt{\frac{8k_BT}{\pi m_2}}e^{-E_b/k_BT},
\end{equation}
where $R_{\text{CO}_2}=3.23$~\AA is the size of the CO$_2$ molecule retrieved from the van der Waals equation of state, $m_2^{-1}=m_{(\text{CO}_2)_2}^{-1}+m_M^{-1}$ and $E_b$ is the electronic binding energy of the dimer (1.27 kcal/mol~\cite{Dingilian2021}). The termolecular reaction rates $k_3$ and $\hat{k}_3$ were calculated from Eq.~(\ref{eq_rate2}) using the corresponding three-body reduced mass and effective long-range coefficients calculated from the values, as explained in the SI. 


Fig.~\ref{fig6} shows again very good agreement between experimental and calculated (filled black circles) rates. It also shows that in the trimer region (blue shaded area) the model with trimers predicts the experiment better than the one with dimers only (filled blue circles). The dimer model overestimates the actual rates because it neglects dimer evaporation, which artificially inflates the rate. The critical cluster size of these regions has been taken from Ref.~\cite{Dingilian2021}. The experimental conditions are identical and the critical cluster number was extracted by comparing the Gibbs free energy of a cluster with the temperature of the system. 

Within this framework, reproducing the experimental nucleation rate in the trimer region requires Eq.~(\ref{eqJ}), which explicitly includes dimer evaporation. The results obtained using Eq.~(\ref{eqJ}) reproduce the experimental data up to $T\lesssim 65$~K. At even higher temperatures, larger critical clusters will become important and thus trimer evaporation will become more significant. Neglecting trimer evaporation (Eq.~(\ref{eqJ})) again overestimates the reaction rate in the gray shaded area. This section demonstrates how the approach can be extended beyond dimers; i.e., into the region where nucleation barriers become important.

\subsection{Range of applicability}

Because it is based on a capture model, our theoretical approach relies on the long-range tail of the intermolecular interaction potential to determine the reaction dynamics. Hence, the present approach is best suited for low-energy collision scenarios, which for termolecular reactions are those in which the collision energy is below the typical binding energy of the reaction products. Indeed, it has been shown that for a collision energy comparable to the binding energy of the main reaction product (dimer)~\cite{mirahmadi2022three}, the short-range region of the interaction potential starts to affect the reaction dynamics, resulting in deviations from Eq.~(\ref{eq_rate2}). This provides a means to estimate the range of applicability of our approach. Since it depends on the specific system, we use here CO$_2$ as an example. (CO$_2$)$_2$ has a binding energy of approximately 0.03~eV $\approx350$~K, which means that reactions with collision energies similar to 350 K depend on short-range details of the interaction potential. At a temperature of about 200~K, such collision energies occur with non-negligible probability. Therefore, we estimate for CO$_2$ that our approach performs well at $T\lesssim 200$~K. For water with higher binding energies this value lies higher at about 500~K due to its larger binding energy~\cite{waterdimer}. 

In the high-pressure regime, our approach will not show the typical plateau that is characteristic of termolecular reaction rates~\cite{Burke}. The reason is that at high pressures, the interparticle distance is of the same order as the LeRoy radius -- the distance at which the electronic clouds of the reactants start to overlap. At this point, electron exchange and correlation effects begin to influence the potential energy landscape. Therefore, we estimate that the present approach is applicable up to total pressures of 0.1~bar or lower. These estimates show that our straightforward approach has a broad applicability range, which can hopefully be tested in the future when more molecular-level data become available. In addition, our model treats molecules as superatoms; i.e., it does not account for internal states in the reaction rate. This is a reasonable approximation for small molecules. However, for larger molecules where internal excitation energies are small, it will be necessary to include the effect of internal degrees of freedom. To this end, one could use the statistical adiabatic channel model from Ref.~\cite{JPR_2024} to estimate the termolecular reaction rate. However, it will be necessary to account for thermochemical information about the reaction partners to calculate the partition function, as is customary in statistical models of reaction dynamics.

\section{Conclusions}

We have established a molecular framework for gas-phase nucleation in which cluster formation emerges from a network of elementary termolecular reactions governed by long-range intermolecular forces. In this picture, the earliest steps of nucleation are not described solely as phenomenological barrier-crossing events, but as direct chemical processes involving nucleating and background species. In single-component mixtures, nucleation proceeds predominantly through the unary channel, where dimers are formed by direct termolecular recombination. In binary mixtures, an additional chaperon channel can accelerate nucleation, with the second species acting as a catalyst for cluster formation.

We tested the theory against unary and binary nucleation of water, butane, and toluene, obtaining agreement in absolute rates within one order of magnitude across the explored concentration, temperature, and density ranges. The theory also captures the experimentally observed temperature dependence, supporting the interpretation of gas-phase nucleation in highly supersaturated environments as elementary termolecular chemistry.

The approach relies primarily on long-range intermolecular interactions, requiring only readily accessible molecular properties and avoiding system-specific fitting or costly atomistic simulations. By incorporating cluster evaporation within the same reaction-network framework, the theory can be extended beyond the collision limit to regimes where larger critical clusters and nucleation barriers become relevant. These results establish a direct connection between termolecular reaction dynamics and nucleation phenomena, providing a molecular foundation for gas-phase nucleation and opening a route toward predictive modeling of cluster formation across chemistry and physics.

\section*{Author contributions}
Y.W.: Investigation, Visualization. A.C.: Investigation, Data curation, Formal analysis, Validation, Writing -- review and editing. F.G.: Investigation, Data curation, Writing -- review and editing. R.S.: Conceptualization, Supervision, Writing -- original draft, Writing -- review and editing. J.P.-R.: Conceptualization, Methodology, Supervision, Writing -- original draft, Writing -- review and editing.

\section*{Conflicts of interest}
There are no conflicts to declare.

\section*{Data availability}
All data and code needed to evaluate and reproduce the results in the paper are present in the paper and/or the SI. This study did not generate any new materials.

\section*{Acknowledgements}
R.S., A.C., and F.G. thank Philipp Albrecht (ETH Z\"urich), Markus Steger (ETH Z\"urich), and Bruce Yoder (ETH Z\"urich) for their technical support. Financial support for the experimental work was provided by the Swiss National Science Foundation (SNSF, grant no. 200021-236446). J.P.-R. and Y.W. acknowledge support from the United States Air Force Office of Scientific Research [grant number FA9550-23-1-0202]. J.P.-R. acknowledges support from the U.S. National Science Foundation under CAREER Award No.~2440808.

{\makeatletter
\providecommand*{\mcitethebibliography}{\thebibliography}
\csname @ifundefined\endcsname{endmcitethebibliography}
{\let\endmcitethebibliography\endthebibliography}{}

\makeatother}


\clearpage
\newgeometry{left=2cm,right=2cm,top=2.3cm,bottom=2.0cm}
\onecolumn
\setlength{\headwidth}{\textwidth}
\setlength{\columnwidth}{\textwidth}
\renewcommand\normalsize{\fontsize{11pt}{13.6pt}\selectfont}%
\normalsize

\setcounter{section}{0}\setcounter{subsection}{0}
\setcounter{figure}{0}\setcounter{table}{0}
\setcounter{equation}{0}\setcounter{page}{1}
\renewcommand{\thepage}{S\arabic{page}}
\renewcommand{\thesection}{S\arabic{section}}
\renewcommand{\thesubsection}{\thesection.\arabic{subsection}}
\renewcommand{\theequation}{S\arabic{equation}}
\renewcommand{\thefigure}{S\arabic{figure}}
\renewcommand{\thetable}{S\arabic{table}}
\renewcommand{\figurename}{Fig.}
\renewcommand{\refname}{Notes and references}

\fancyhf{}
\fancyhead[L]{\itshape\footnotesize Journal of the Royal Society of Chemistry}
\fancyhead[R]{\itshape\footnotesize Electronic Supplementary Information}
\fancyfoot[C]{\footnotesize \thepage}
\renewcommand{\headrulewidth}{0.4pt}
\renewcommand{\footrulewidth}{0pt}
\pagestyle{fancy}

\titleformat{\section}{\normalfont\fontsize{12.5pt}{15pt}\bfseries\selectfont}{\thesection}{0.7em}{}
\titleformat{\subsection}{\normalfont\fontsize{11pt}{13pt}\bfseries\selectfont}{\thesubsection}{0.7em}{}
\titlespacing*{\section}{0pt}{12pt}{5pt}
\titlespacing*{\subsection}{0pt}{9pt}{3pt}

\thispagestyle{fancy}
\begin{center}
{\large\bfseries Electronic Supplementary Information (ESI)}\\[10pt]
{\LARGE\bfseries A termolecular reaction theory for gas-phase nucleation based on long-range intermolecular forces}\\[12pt]
{\normalsize Yu Wang, Arnab Choudhury, Felix Graber, Ruth Signorell, Jes\'us P\'erez-R\'ios}\\[8pt]
{\footnotesize $^{\ast}$Corresponding author. Email: \href{mailto:jesus.perezrios@stonybrook.edu}{jesus.perezrios@stonybrook.edu}}
\end{center}
\vspace{6pt}
\noindent\rule{\linewidth}{0.4pt}
\vspace{10pt}

\section{Capture model and termolecular reaction rates}
\begin{figure}[t]
    \centering
    \includegraphics[width=\linewidth]{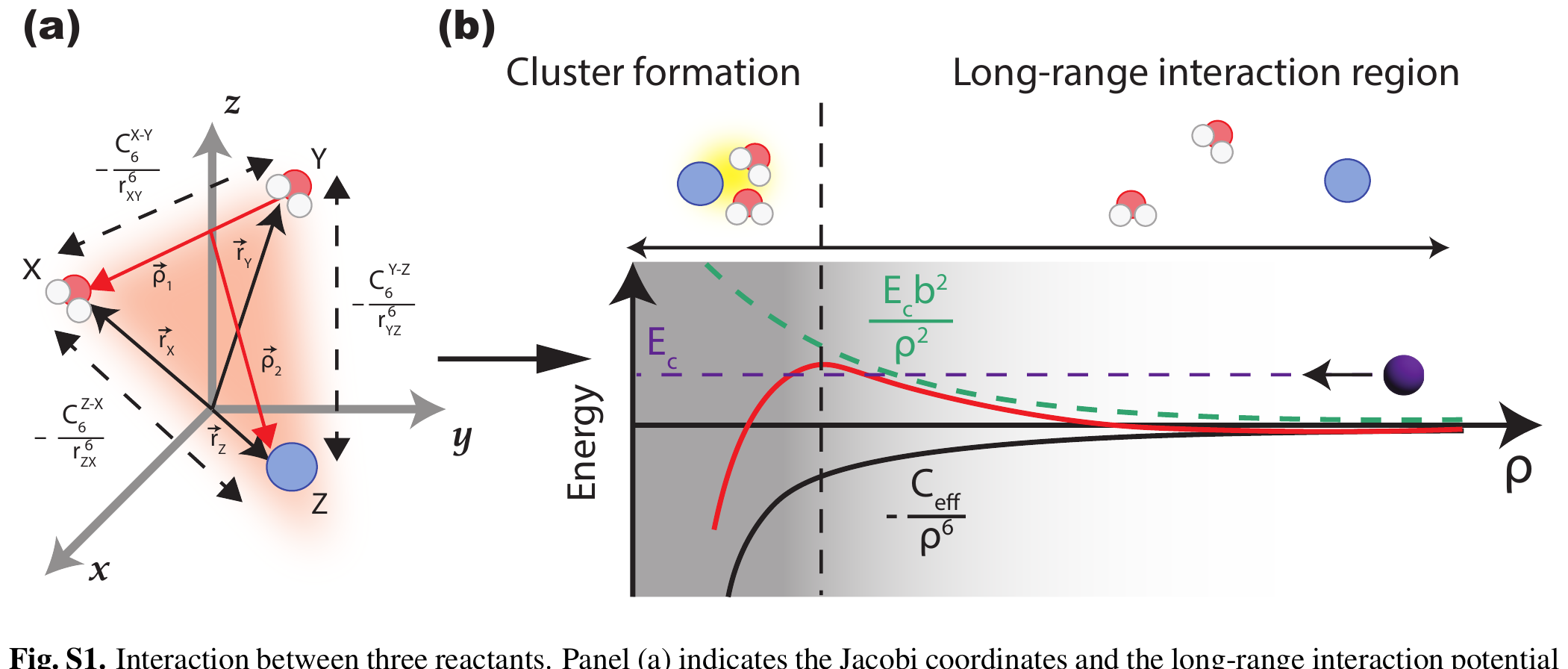}
    \caption{Interaction between three reactants. Panel (a) indicates the Jacobi coordinates and the long-range interaction potential among the reactants. $\vec{\rho}_1$ represents the interparticle distance vector between X and Y reactants, and $\vec{\rho}_2$ describes the motion of the center of mass of the X-Y pair with respect to the third reactant, Z. Panel (b) represents the effective hyperradial potential of the particles of the left panel. Panel (b) shows a rendering of the reaction dynamics using a capture model in hyperspherical coordinates. The height of the effective potential (long-range plus centrifugal barrier) determines the onset of chemistry, so that any incoming state with energy above the potential will lead to dimer formation.}
    \label{fig:interaction}
\end{figure}

For the interaction energy landscape of a termolecular reaction $\mathrm{X}+\mathrm{Y}+\mathrm{Z}$, the long-range interaction reads as
\begin{equation}
\label{eq:V1}
V(\mathbf{r}_\mathrm{X},\mathbf{r}_\mathrm{Y},\mathbf{r}_\mathrm{Z})=
\frac{C_6^{\text{X-Y}}}{r_{XY}^{6}}
+\frac{C_6^{\text{Y-Z}}}{r_{YZ}^{6}}
+\frac{C_6^{\text{Z-X}}}{r_{ZX}^{6}}
\end{equation}
where the $C_6^{\text{X-Y}}$--coefficients denote the long-range van der Waals coefficients between two of the three molecules, X-Y, in this case and $r_{XY}$ is the distance between them. This interaction scheme is shown in Fig.~\ref{fig:interaction}(a). However, to define the capture radius, a reaction coordinate must be specified. To this end, it is better to use Jacobi coordinates, given by
\begin{equation}
\label{eq:jacobi}
\begin{aligned}
&\vec{\rho}_1=\vec{r}_\mathrm{Y}-\vec{r}_\mathrm{X}\,,\qquad
\vec{\rho}_2=\vec{r}_\mathrm{Z}-\vec{R}_{XY}\,,\\[4pt]
&\vec{\rho}_{CM}=\frac{m_\mathrm{X}\vec{r}_\mathrm{X}+m_\mathrm{Y}\vec{r}_\mathrm{Y}+m_\mathrm{Z}\vec{r}_\mathrm{Z}}{M}\,,
\end{aligned}
\end{equation}
where $M=m_\mathrm{X}+m_\mathrm{Y}+m_\mathrm{Z}$ is the total mass and $\vec{R}_{XY}=(m_\mathrm{X}\vec{r}_\mathrm{X}+m_\mathrm{Y}\vec{r}_\mathrm{Y})/(m_\mathrm{X}+m_\mathrm{Y})$ and the mass of each of the molecules is denoted by $m_\mathrm{X}$, $m_\mathrm{Y}$ and $m_\mathrm{Z}$, respectively. Accounting solely for the translational degrees of freedom, it is preferable to describe the configuration of the three reactants as\cite{SI:PerezRios2014}
\begin{equation}
\label{eq:rho6d}
\vec{\rho}_{6D}=\begin{pmatrix}\rho_1\\[2pt]\rho_2\end{pmatrix},
\end{equation}
that we describe in hyperspherical coordinates. Hyperspherical coordinates can be viewed as the generalization of spherical coordinates to a 6-dimensional sphere, including one hyperradius $\rho$ and 5 hyperangles $\alpha=(\alpha_1,\alpha_2,\alpha_3,\alpha_4,\alpha_5)$\cite{SI:Avery}. Thereby, the long-range interaction potential given by Eq.~(\ref{eq:V1}) reads as\cite{SI:Mirahmadi}
\begin{equation}
\label{eq:Vrhoalpha}
V(\rho,\alpha)=-\frac{C_6^{X-Y}}{r_{XY}(\rho,\alpha)^{6}}
-\frac{C_6^{Y-Z}}{r_{YZ}(\rho,\alpha)^{6}}
-\frac{C_6^{Z-X}}{r_{ZX}(\rho,\alpha)^{6}},
\end{equation}
explicitly showing that the interparticle distance is a function of the hyperradius and hyperangles.

The potential~(\ref{eq:Vrhoalpha}) can be further simplified by assuming the adiabatic approximation, such that the angular degrees of freedom are faster than the radial ones. Therefore, it is possible to integrate the hyperangular degrees of freedom out of the interaction potential, yielding an effective interaction potential as shown in Fig.~\ref{fig:interaction}(b), defined as\cite{SI:Mirahmadi,SI:YuWang}
\begin{equation}
\label{eq:Veff}
V(\rho)=-\frac{C_6^{\mathrm{eff}}}{\rho^{6}}.
\end{equation}

In this equation, $C_{eff}$ is the effective long-range coefficient for the termolecular reaction, or the effective three-body van der Waals coefficient. Therefore, the hyperradius is considered the reaction coordinate. Hence, for a given collision energy, it is possible to define the capture hyperradius that separates elastic collisions from chemical reactions, as shown in panel (b) of Fig.~\ref{fig:interaction}. The capture hyperradius is determined as the maximum impact parameter $b(E_c)_{\max}$ for a given collision energy $E_c$, for which reactants can visit the short-range region where cluster formation occurs. Assuming that when reactants visit the cluster formation region, represented as the vertical dashed line in panel (b), the reaction occurs with 100\% efficiency, the recombination cross section is given by
\begin{equation}
\label{eq:sigma3}
\sigma_3(E_c)=\frac{8\pi^2}{3}\int_0^{b(E_c)_{\max}}b^4\,db=\frac{8\pi^2}{15}b^5(E_c)_{\max},
\end{equation}
and after performing the Maxwell-Boltzmann average, we get the recombination rate\cite{SI:Mirahmadi}
\begin{equation}
\label{eq:k3}
k_3(T)=\frac{4\pi^3}{3\Gamma(1/3)\sqrt{\mu_3}}\left(2C_6^{\mathrm{eff}}\right)^{5/6}(k_BT)^{-1/3},
\end{equation}
where $\mu_3=\sqrt{m_\mathrm{X}m_\mathrm{Y}m_\mathrm{Z}/(m_\mathrm{X}+m_\mathrm{Y}+m_\mathrm{Z})}$ is the three-body reduced mass, $k_B$ the Boltzmann constant, $T$ the temperature, and $\Gamma$ the Gamma function. It is worth noting that Eq.~(\ref{eq:k3}) shows a negative power-law behavior with respect to the temperature, as expected for barrier-less reactions such as recombination reactions.

\section{Pair-wise van der Waals coefficients}
To evaluate the effective long-range coefficient for any termolecular reaction, we first need the pair-wise van der Waals coefficients for all relevant interactions. For the systems considered in this work, the systems are A-A, A-B, A-M, and M-M, where A is butane, toluene, or water; M is N$_2$ or Ar; and B = CO$_2$. The values of the pair-wise van der Waals coefficients are shown in Table~\ref{tab:pairwise}. Some of them are not experimentally available, or no ab initio quantum-chemical calculation is available. Those cases were determined from the London dispersion relation, which for the van der Waals coefficients reads as\cite{SI:Tkatchenko}
\begin{equation}
\label{eq:london}
C_6^{A-B}=\frac{2C_6^{A-A}C_6^{B-B}}{\dfrac{\alpha_B}{\alpha_A}C_6^{A-A}+\dfrac{\alpha_A}{\alpha_B}C_6^{B-B}},
\end{equation}
where $\alpha_Z$ denotes the static polarizability of the Z species, taken from NIST. The specific values of the static polarizability of the relevant species are shown in Table~\ref{tab:polar}.

\begin{table}[h!]
\begin{center}
\caption{Pair-wise van der Waals coefficients of the nucleating species, catalysts, and carrier gases involved in nucleation. All the coefficients are expressed in atomic units (Hartree $\times$a$_0^6$, being a$_0$ the Bohr radius.)}\label{tab:pairwise}
\begin{tabular}{lc}\toprule
System& Value \\
\midrule
C$_4$H$_{10}$-C$_4$H$_{10}$  & 1785$^\text{a}$\\
C$_4$H$_{10}$-Ar  & 334\\
C$_4$H$_{10}$-N$_2$  & 358\\
C$_4$H$_{10}$-CO$_2$  & 578\\
\hline
C$_7$H$_{8}$-C$_7$H$_{8}$ &   2625$^\text{b}$\\
C$_7$H$_{8}$-Ar &   408$^\text{b}$\\
C$_7$H$_{8}$-N$_2$ &   435$^\text{b}$\\
C$_7$H$_{8}$-CO$_2$ &   639$^\text{b}$\\
\hline
H$_2$O-H$_2$O & 45.4$^\text{c}$\\
H$_2$O-Ar & 54\\
H$_2$O-N$_2$ & 58\\
H$_2$O-CO$_2$ & 92\\
\hline
CO$_2$-CO$_2$ & 192.0$^\text{d}$\\
N$_2$-N$_2$ & 73.4$^\text{e}$\\
Ar-Ar & 64.2$^\text{f}$\\
CO$_2$-N$_2$ & 118.2\\
CO$_2$-Ar & 110.15\\
\hline
(CO$_2$)$_2$-CO$_2$ & 230.4$^\text{h}$\\
(CO$_2$)$_2$-Ar& 132.18$^\text{h}$\\
\bottomrule
\end{tabular}
\end{center}
\footnotesize{\textsf{[a] Value estimated from the van der Waals equation of state of butane. [b] Value taken from Ref.~\cite{SI:Toluene}. [c] Value taken from Ref.~\cite{SI:water}. [d] Value taken from Ref.~\cite{SI:CO2}. [e] Value taken from Ref.~\cite{SI:N2}. [f] Value taken from Ref.~\cite{SI:Ar}. [g] Value estimated as 20\% larger than for the equivalent (CO$_2$)-X long range coefficient.}}
\end{table}

\begin{table}[h]
\begin{center}
\caption{Polarizability of the chemical species relevant for this study (in atomic units)}\label{tab:polar}
\begin{tabular}{lc}\toprule
Molecule& Polarizability (atomic units) \\
\midrule
C$_4$H$_{10}$& 54.12 \\
C$_7$H$_{8}$ & 80.04 \\
H$_2$O &9.65 \\
Ar & 11.1\\
N$_2$ & 11.54\\
CO$_2$ & 16.92\\
\bottomrule
\end{tabular}
\end{center}
\end{table}

\section{Effective long-range coefficient for termolecular reactions}
To evaluate the effective long-range coefficient for any termolecular reaction, we first need the pairwise van der Waals coefficients for all relevant molecule-molecule and molecule-atom interactions. For the systems considered here, these are A-A, A-B, A-M, and M-M, where A is butane, toluene, or water, M is N$_2$ and Ar, and B = CO$_2$, and their values are reported in Table~\ref{tab:pairwise}. With this information at hand, we use Eq.~(\ref{eq:Vrhoalpha}) to generate different angular configurations. For each of these configurations, we find the value of the hyperradius that solves $V(\rho^{*},\alpha)=E_c$, such that $\rho^{*}$ is the capture hyperradius. For a large number of angular configurations, we obtain a distribution of capture hyperradius, and we pick the most probable hyperradius as the capture hyperradius $\rho$ for a given collision energy $E_c$. By repeating this procedure for a set of different collision energies, we find the effective long-range interaction coefficient by solving for $-C_6^{\mathrm{eff}}/\rho^6=E_c$, and the results are shown in Table~\ref{tab:ceff3body}. These values are fed into Eq.~(\ref{eq:k3}) to calculate the reaction rates for all relevant termolecular reactions. It is worth noting that a more detailed discussion of the procedure can be found in Ref.~\cite{SI:Mirahmadi}.

\begin{table}[h]
\centering
\caption{Effective three-body long range coefficients for all vapor components considered in this work. All coefficients are expressed in atomic units (Hartree $\times$a$_0^6$, being a$_0$ the Bohr radius.)}
\label{tab:ceff3body}
\vspace{1mm}
\adjustbox{max width=\linewidth}{%
\begin{tabular}{llcccc}\toprule
A & X & A-A-X & A-CO$_2$-X & A-X-X & A-Ar-N$_2$\\
\midrule
\multirow{3}{*}{H$_2$O} & H$_2$O & --- & 2248 & 1305 & 1763\\
                        & Ar     & 1459 & 2701 & 1703 & ---\\
                        & N$_2$  & 1558 & 2705 & 1847 & ---\\
\midrule
\multirow{3}{*}{C$_4$H$_{10}$} & C$_4$H$_{10}$ & ---   & 26854 & 51349 & 7194\\
                               & Ar            & 21032 & 9329  & 6642  & ---\\
                               & N$_2$         & 24931 & 10368 & 7622  & ---\\
\midrule
\multirow{3}{*}{C$_7$H$_{8}$} & C$_7$H$_{8}$ & ---   & 39868 & 75391 & 9050\\
                              & Ar           & 34946 & 11180 & 8488  & ---\\
                              & N$_2$        & 43257 & 12333 & 9342  & ---\\
\bottomrule
\end{tabular}}
\end{table}

\subsection{Robust extraction of the capture hyperradius}
As described above, $C_6^{\mathrm{eff}}$ is obtained as the slope of $E_c$ versus $1/\varrho^6$, where $\varrho$ is the mode of the hyperradius distribution at each collision energy $E_c$. Due to the sharp lower bound of the distribution it is necessary to design a robust method to extract the mode of the distribution. Here, we propose to use a kernel density estimate (KDE) with Gaussian kernel to analyze the distribution of hyperradius. KDE produces a continuous, parameter-stable estimate of the density and is particularly suited to the dense bulk of our sampled $\varrho$ distribution, where histogram-based modes are sensitive to the choice of bin width. To asses the robustness of the approach we use bootstrap resampling. We find that the KDE mode exhibits a 95\% confidence interval whose width is comparable to the errorbar of $\rho$ samples, propagated from the uncertainty of pair-wise $C_6$, indicating that the estimate is robust to statistical resampling.

The hyperradius distribution also admits a parametric description. Previous work has established the universal tail behavior of the capture hyperradius: the upper tail lies in the maximum domain of attraction of the Fr\'echet distribution, a consequence of the local volume measure on $\mathcal{S}^5$ together with the inverse-sixth interaction. We have therefore fitted the full distribution to the Generalized Extreme Value (GEV) family in previous work and extracted $C_6^{\mathrm{eff}}$ from the GEV mode. We note that the GEV provides a rigorous description of the tail. However, the GEV characterization applies primarily to the tail, and deviations from a pure GEV form are expected in the bulk of the KDE-sampled distribution. However, this deviation is within the 30\% error claimed in the main text.

\section{Details on unary nucleation}
\begin{figure}[h]
    \centering
    \includegraphics[width=\linewidth]{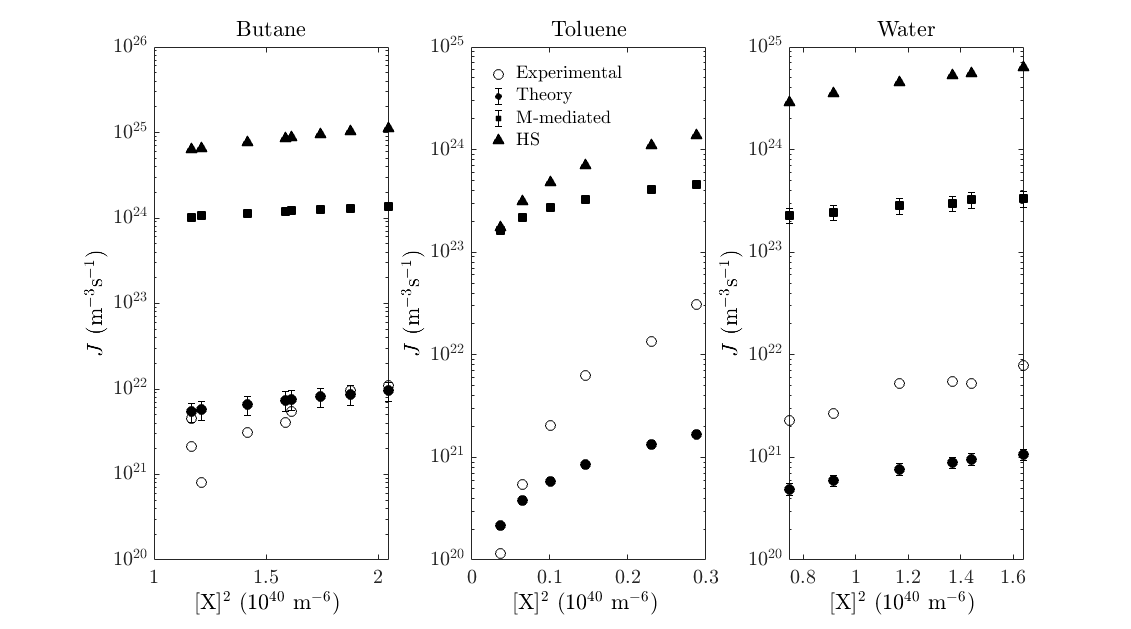}
    \caption{Unary nucleation rates for butane, toluene, and water as a function of the square concentration of the nucleating vapor for 51, 55, and 57~K, respectively. The empty symbols represent the experimental results; the filled circles show the results from our theory; the squares represent the results from the M-mediated reaction channel; and the filled triangles depict the hard-sphere model predictions (HS). See the text for the errorbars.}
    \label{fig:unary}
\end{figure}

In principle, the inert gas could also react with any of the other two reactants to form a van der Waals complex, as it is observed in buffer gas sources or jets~\cite{SI:Tariq2013,SI:Quiros2017,SI:Brahms2011,SI:Koperski2002}, via the reaction
\begin{equation}
\label{r4}
\ce{A + M + M ->[$k^{M}_{3,M}$] AM + M}.
\end{equation}
In our work, we refer to this reaction channel as the M-mediated channel. In general, AM will be a weakly bound molecule with a binding energy $\sim 1$~meV, that can undergo a second bimolecular reaction
\begin{equation}
\label{r5}
\ce{AM + A ->[$k^{M}_2$] A2 + M}
\end{equation}
resulting in a dimer. Reaction~(\ref{r5}) is highly efficient since the binding energy of the dimer is larger than the AM complex, so the rate-determinant process is indeed reaction~(\ref{r4}). Therefore, the contribution to the nucleation rate of the dimer through the M-mediated channel is given by
\begin{equation}
\label{M-mediated}
J_{\text{M-mediated}}=k^M_{3,M}[M][M][A],
\end{equation}
and the results are shown by the squares in Fig.~\ref{fig:unary}. The M-mediated reaction channel overestimates the unary nucleation rate by a factor of $\sim 50$; hence, we conclude that they do not play a role in nucleation. The carrier gas acts as an expectant third body during the nucleation process, so it does not accelerate or affect the chemical kinetics of nucleation. Furthermore, we observe that the HS prediction is even larger than when the M-mediated reaction channel is included, indicating that the HS model systematically overestimates the unary nucleation rates.

\subsection{Hard sphere model}
In the hard-sphere model of reaction rates, it is assumed that the reaction occurs with 100\% probability beyond a threshold, $R$, the molecular size. In this case, the two-body reaction rate is given by
\begin{equation}
\label{Eq_HS}
k_{HS}=\sigma_{HS}\langle v\ \rangle=4R^2\sqrt{\frac{16\pi k_BT}{m}},
\end{equation}
where $m$ is the mass of the nucleating species, and $T$ is the temperature of the system. Here, we estimate the molecular size via the van der Waals equation of state. Specifically, from the value of the parameter $b$ in the van der Waals equation of state. The nucleation rate is finally calculated as
\begin{equation}
\label{Eq_HS2}
J=k_{HS}\rho^2,
\end{equation}
where $\rho$ is the number density of the nucleating species.

\begin{figure}[h]
    \centering
    \includegraphics[width=0.5\linewidth]{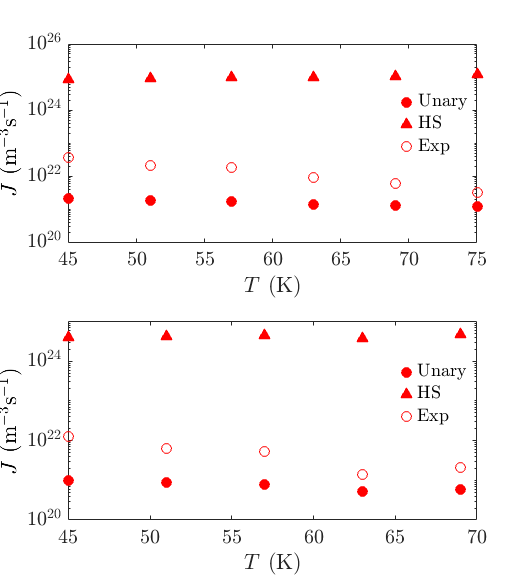}
    \caption{Temperature-dependent water nucleation rate for a given fixed water density. The experimental data for each panel were taken keeping the water density as close as possible to $1.60\times 10^{20}$~m$^{-3}$ and $1.07\times 10^{20}$~m$^{-3}$ for panels (a) and (b), respectively. The unary and hard-sphere (HS) results are calculated for the specific water density in each panel.}
    \label{fig:thermal}
\end{figure}

It is very important to realize that Eq.~(\ref{Eq_HS2}) shows a positive temperature dependence: at higher temperatures, the nucleation rate is much larger. This is clearly a flaw in the formulation, since nucleation in the barrierless case should be inversely proportional to temperature, as shown in the experimental data in Fig.~\ref{fig:thermal}. In this figure, we see that the HS model does not capture the temperature dependence of water's nucleation rate, whereas our theoretical results better match the experimental results.

\subsection{Theoretical errorbars}
The results of our model depend on the value of the effective long-range coefficient. Those depend primarily on the values of the pairwise van der Waals coefficients among the molecular partners and, to a lesser extent, on the choice of Jacobi coordinates (the labels of the molecules for the choice of coordinates).

In this work, we estimate the van der Waals coefficient from a van der Waals equation of state for butane, which introduces some error ($\sim$30\%) that can propagate when calculating the effective termolecular long-range coefficient. Therefore, for butane, we include an error bar of 30\% to account for this. In the case of water, some of the pairwise van der Waals coefficients have been obtained via Eq.~(\ref{eq:london}), which is only a proxy for the real van der Waals coefficient, so we assume an error of 15\% for water nucleation rates. Finally, for toluene, we have accurate inter-particle van der Waals coefficients~\cite{SI:Toluene}, so the effective long-range coefficient is more accurate than in the other two systems considered. However, we need to account for fluctuations in the effective van der Waals coefficient when using different Jacobi coordinates, leading to an error of less than 10\%. Therefore, the toluene rates have an error bar of 10\%.

For the unary calculations for the regime where the critical cluster is larger than a dimer, the errorbars are considered to be 30\% for $k_3$ and 50\% for $k_3'$ since the last one is estimated contemplating size effects of (CO$_2$)$_2$ with respect CO$_2$. The termolecular reaction rates $k_3$ and $\hat{k}_3$ were calculated from Eq.~(15) of the main text, using the corresponding three-body reduced mass and effective long range coefficients calculated from Table~\ref{tab:pairwise}. The values for $(\text{CO}_2)_2$-X were estimated as 20\% larger than those for CO$_2$-X interactions. This approximation can induce errors as large as 50\%, which are accounted for in the error bars shown in Fig.~7 of the main text.

\section{Experimental setup}
\begin{figure}[h]
    \centering
    \includegraphics[width=\linewidth]{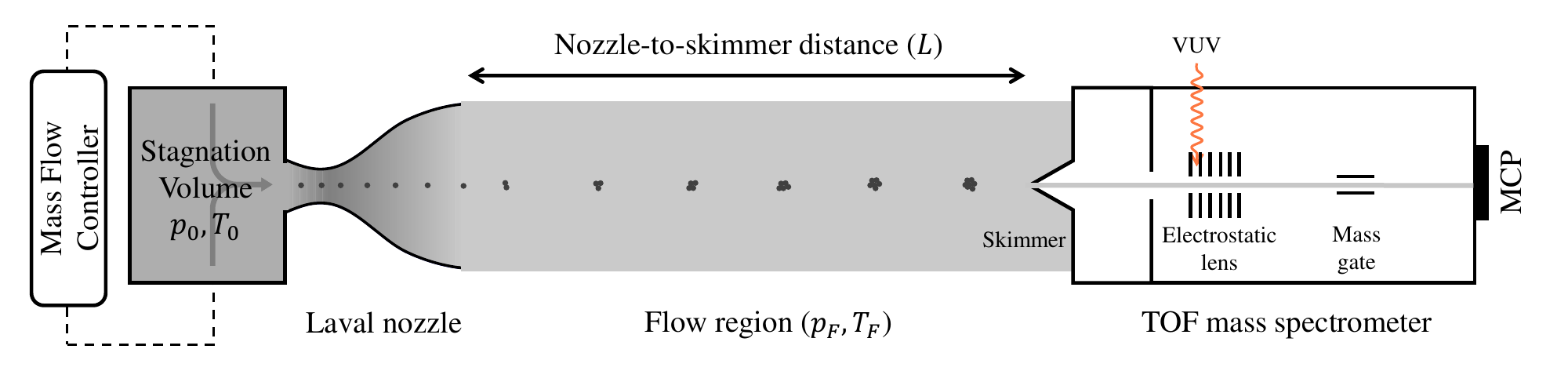}
    \caption{Schematic of the Laval experimental setup. Nucleation takes place in the post-nozzle flow of the Laval expansion. The formed clusters are detected in the time-of-flight (TOF) mass spectrometer after soft photoionization. $p_0$ and $T_0$ are the pressure and temperature in the stagnation volume and $p_F$ and $T_F$ are the pressure and temperature of the flow. The nucleation time ($t$) can be varied by varying the nozzle-to-skimmer distance ($L$). The figure is reused from Ref~\cite{SI:Choudhury2026} with permission from the PCCP Owner Societies.}
    \label{SIfig:exp_setup}
\end{figure}

The experiments were carried out in our Laval setup depicted in Figure~\ref{SIfig:exp_setup}. This section is only an overview, for a more detailed version see~\cite{SI:Ruth_1,SI:Li_2019}. A gas mixture consisting of carrier gas (argon and nitrogen), a mass spectrometry standard (methan) as well as the nucleating species (butane, water or toluene) are expanded through a laval nozzle. For our binary experiments, parts of the carrier gas were replaced by CO$_2$. Unless otherwise mentioned, all experiments were performed using a Mach 4 Laval nozzle at 40 Pa chamber pressure. The nucleating species can then nucleate in the uniform flow after the Laval nozzle. We measure static pressure, gas velocity, and temperature in this post-nozzle flow using a pitot tube~\cite{SI:Ruth_1}, which gives us information about the nucleation conditions. The Laval nozzle is mounted on a translation stage, so that we can freely change the length of the post nozzle flow, allowing us to probe different nucleation times.

The clusters which are formed in the post-nozzle flow are then analyzed in a home built mass spectrometer. They are softly ionized using single photon ionization with a photon energy slightly above the first ionization energy. The cluster ions are accelerated in an electric field towards a micro channel plate detector, where the clusters are detected, providing information on their chemical composition and number concentration.

The supersaturation of the nucleating species was calculated using the Wagner equation and coefficients from Ref.~\cite{SI:VDI} for the equilibrium vapor pressures. The soft single-photon ionization was performed with a home-built vacuum-ultraviolet (VUV) laser at a photon energy of 13.8~eV, just above the lowest ionization energy. The conditions were chosen so that the concentration of the carrier gas (M) exceeds that of the condensable gas (A) by several hundred to several thousand times. Furthermore, the catalyst B (here B = CO$_2$) was also present in excess compared to the condensable gas, typically exceeding the concentration of A by up to a few tens of times; the catalyst cannot condense under these conditions. The specific conditions and nucleation rates for the different experiments are listed in the data tables of Section~\ref{sec:data}.

\clearpage
\section{Data overview}\label{sec:data}
\subsection{Unary nucleation}
This section contains data for unary nucleation of butane (Table~\ref{SItab:unary_butane_data}), toluene (Table~\ref{SItab:unary_toluene_data}), and water (Table~\ref{SItab:unary_water_data}). All experiments were performed at a chamber pressure of 40 Pa. The column contains: the concentration of the nucleating species ($C_{nu}$) [$nu$: butane ($Bu$), toluene ($Tol$), and water ($H_2O$)], natural logarithm of supersaturation ($\ln(\mathrm{S})$), nitrogen concentration ($C_{N_2}$), argon concentration ($C_{Ar}$), the square of the concentration of teh nucleating species ($C_{nu}^2$), experimental nucleation rate ($J_{exp}$) with its error ($J^{\text{err}}_{exp}$), the theoretical nucleation rate according to 3-body recombination approach ($J_{3BR}$) with its error ($J^{\text{err}}_{3BR}$), and the hard-sphere reaction rate ($J_{HS}$).

\begin{table}[h]
    \centering
    \caption{Overview table for unary butane nucleation measurements at flow temperature 51 K. The experimental data is taken from Ref~\cite{SI:Choudhury2026}.}
    \vspace{3mm}
    \label{SItab:unary_butane_data}
    \adjustbox{max width=\linewidth}{%
\begin{tabular}{|c|c|c|c|c|c|c|c|c|c|}\hline\hline
         $C_{Bu}$&  $\ln(S)$&  $C_{N_2}$&  $C_{Ar}$&  $C_{Bu}^2$&  $J_{exp}$&  $J^{\text{err}}_{exp}$&  $J_{3BR}$&  $J^{\text{err}}_{3BR}$& $J_{HS}$
\\
         $(10^{20}$ m$^{-3})$&  &  $(10^{21}$ m$^{-3})$&  $(10^{22}$ m$^{-3})$&  $(10^{40}$&  $(10^{21}$&  $(10^{21}$&  $(10^{21}$&  $(10^{21}$& $(10^{24}$
\\
         &  &  &  &  m$^{-6})$&  m$^{-3}$s$^{-1})$&  m$^{-3}$s$^{-1})$&  m$^{-3}$s$^{-1})$&  m$^{-3}$s$^{-1})$& m$^{-3}$s$^{-1})$
\\\hline
         1.08&  45&  7.16&  4.95&  1.17&  4.5&  0.4&  5.4&  1.4& 6.3\\
         1.10&  46&  7.27&  5.03&  1.21&  0.8&  0.2&  5.7&  1.5& 6.6\\
         1.08&  44&  7.13&  4.94&  1.17&  2.1&  0.3&  5.4&  1.4& 6.3\\
         1.19&  44&  7.12&  4.94&  1.42&  3.1&  0.5&  6.6&  1.7& 7.7\\
         1.26&  45&  7.16&  4.97&  1.59&  4.1&  0.3&  7.4&  1.9& 8.6\\
         1.27&  46&  7.22&  5.01&  1.61&  5.4&  0.6&  7.6&  1.9& 8.7\\
         1.32&  45&  7.16&  4.97&  1.74&  8.1&  0.8&  8.1&  2.1& 9.4\\
         1.37&  45&  7.11&  4.94&  1.88&  9.5&  1.2&  8.7&  2.2& 10.2\\
         1.43&  45&  7.15&  4.97&  2.04&  11.0&  1.6&  9.5&  2.5& 11.1\\\hline\hline
    \end{tabular}}
\end{table}

\begin{table}[h]
    \centering
   \caption{Overview table for unary toluene nucleation measurements at flow temperature 55 K. The experimental data is newly measured and has not been published before.}
    \vspace{3mm}
    \label{SItab:unary_toluene_data}
    \adjustbox{max width=\linewidth}{%
\begin{tabular}{|c|c|c|c|c|c|c|c|c|c|}\hline\hline
         $C_{Tol}$&  ln(S)&  $C_{N_2}$&  $C_{Ar}$&  $C_{Tol}^2$&  $J_{exp}$&  $J^{\text{err}}_{exp}$&  $J_{3BR}$&  $J^{\text{err}}_{3BR}$& $J_{HS}$
\\
         $(10^{19}$ m$^{-3})$&  &  $(10^{22}$ m$^{-3})$&  $(10^{22}$ m$^{-3})$&  $(10^{39}$&  $(10^{20}$&  $(10^{20}$&  $(10^{20}$&  $(10^{20}$& $(10^{23}$
\\
         &  &  &  &  m$^{-6})$&  m$^{-3}$s$^{-1})$&  m$^{-3}$s$^{-1})$&  m$^{-3}$s$^{-1})$&  m$^{-3}$s$^{-1})$& m$^{-3}$s$^{-1})$
\\\hline
         1.92&  72&  1.48&  3.79&  0.37&  1.2&  0.3&  2.2&  0.2& 1.8\\
         2.56&  71&  1.48&  3.77&  0.66&  5.5&  0.6&  3.8&  0.3& 3.1\\
         3.17&  72&  1.48&  3.79&  1.00&  20.4&  2.4&  5.9&  0.5& 4.8\\
         3.82&  72&  1.48&  3.77&  1.46&  63.4&  9.0&  8.5&  0.7& 6.9\\
         4.80&  72&  1.48&  3.78&  2.30&  134.0&  8.4&  13.5&  1.1& 11.0\\
         5.37&  73&  1.48&  3.79&  2.88&  306.0&  36.9&  16.9&  1.4& 13.7\\ \hline\hline
    \end{tabular}}
\end{table}

\begin{table}[h]
    \centering
    \caption{Overview table for unary water nucleation measurements at flow temperature 57 K. The experimental data is remeasured and reproduced version of ref~\cite{SI:Feusi2024}.}
    \vspace{3mm}
    \label{SItab:unary_water_data}
    \adjustbox{max width=\linewidth}{%
\begin{tabular}{|c|c|c|c|c|c|c|c|c|c|}\hline\hline
         $C_{H_2O}$&  ln(S)&  $C_{N_2}$&  $C_{Ar}$&  $C_{H_2O}^2$&  $J_{exp}$&  $J^{\text{err}}_{exp}$&  $J_{3BR}$&  $J^{\text{err}}_{3BR}$& $J_{HS}$
\\
         $(10^{20}$ m$^{-3})$&  &  $(10^{22}$ m$^{-3})$&  $(10^{22}$ m$^{-3})$&  $(10^{40}$&  $(10^{21}$&  $(10^{21}$&  $(10^{20}$&  $(10^{20}$& $(10^{24}$
\\
         &  &  &  &  m$^{-6})$&  m$^{-3}$s$^{-1})$&  m$^{-3}$s$^{-1})$&  m$^{-3}$s$^{-1})$&  m$^{-3}$s$^{-1})$& m$^{-3}$s$^{-1})$
\\\hline
         0.86&  80&  1.77&  3.34&  0.74&  2.3&  0.2&  4.9&  0.6& 2.9\\
         0.96&  78&  1.74&  3.28&  0.91&  2.7&  0.3&  5.9&  0.7& 3.5\\
         1.08&  80&  1.77&  3.33&  1.17&  5.3&  0.3&  7.7&  1.0& 4.5\\
         1.20&  82&  1.79&  3.38&  1.44&  5.2&  0.4&  9.6&  1.2& 5.5\\
         1.17&  78&  1.74&  3.28&  1.37&  5.5&  0.5&  8.9&  1.1& 5.3\\
         1.28&  79&  1.76&  3.31&  1.64&  7.8&  0.5&  10.7&  1.4& 6.3\\ \hline\hline
    \end{tabular}}
\end{table}

\clearpage
\subsection{Temperature dependence of nucleation rates}
This section contains the temperature dependent nucleation rates of water at a chamber pressure of 40 Pa and at a particular water monomer concentration. The columns contain: temperature ($T$), the water concentration ($C_{H_2O}$), natural logarithm of supersaturation ($\ln(\mathrm{S})$), argon concentration ($C_{Ar}$), square of the water concentration ($C_{H_2O}^2$), nitrogen concentration ($C_{N_2}$), experimental nucleation rate ($J_{exp}$) with its error ($J^{\text{err}}_{exp}$), the theoretical nucleation rate according to 3-body recombination approach ($J_{3BR}$), and the hard-sphere reaction rate ($J_{HS}$).

\begin{table}[H]
    \centering
    \caption{Overview table for temperature study of unary water nucleation at water concentration $\approx$ 1.3$\times$10$^{20}$  m$^{-3}$. All data shown here has not been published before.}
    \vspace{3mm}
    \label{SItab:temp1_water_data}
    \adjustbox{max width=\linewidth}{%
\begin{tabular}{|c|c|c|c|c|c|c|c|c|}  \hline\hline
          $T$&$C_{H_2O}$&  ln(S)&  $C_{Ar}$&  $C_{N_2}$&  $J_{exp}$&  $J^{\text{err}}_{exp}$&  $J_{3BR}$& $J_{HS}$
\\
          $(K)$&$(10^{20}$ m$^{-3})$&  &  $(10^{22}$ m$^{-3})$&  $(10^{22}$ m$^{-3})$&  $(10^{21}$&  $(10^{21}$&  $(10^{20}$& $(10^{24}$
\\
          &&  &  &  &  m$^{-3}$s$^{-1})$&  m$^{-3}$s$^{-1})$&  m$^{-3}$s$^{-1})$& m$^{-3}$s$^{-1})$
\\ \hline
          45&1.34&  108&  6.33&  0&  25.5&  1.2&  15.3& 6.1\\
          51&1.33&  94&  4.84&  0.88&  12.6&  1.5&  13.3& 6.4\\
          57&1.38&  78&  3.27&  1.74&  10.4&  0.8&  12.3& 7.3\\
          63&1.36&  69&  2.27&  2.34&  5.3&  0.4&  10.8& 7.5\\
          69&1.33&  60&  1.36&  2.84&  3.1&  0.3&  9.3& 7.5\\ \hline\hline
    \end{tabular}}
\end{table}

\clearpage
\subsection{Binary nucleation with water}
This section contains the comparison between the experimental and theoretical nucleation rates for binary nucleation of butane (Table~\ref{SItab:butane_binary}), toluene (Table~\ref{SItab:tol_binary}), and water (Table~\ref{SItab:water_binary}) mixed with \ce{CO2}. The column contains: the concentration ratio between \ce{CO2} and the nucleative substance ($C_{CO_2}/C_{nu}$) [$nu$: butane ($Bu$), toluene ($Tol$), and water ($H_2O$)], experimental nucleation rate ($J_{exp}$), theoretical nucleation rate considering all channels using 3-body recombination approach ($J_{3BR}$) with its error ($J^{\text{err}}_{3BR}$), theoretical nucleation rate through Chaperon mechanism ($J_{\text{Chp}}$) with its error ($J^{\text{err}}_{\text{Chp}}$), theoretical unary nucleation rate ($J_{\text{un}}$) with its error ($J^{\text{err}}_{\text{un}}$) and the hard-sphere reaction rate ($J_{HS}$).

\begin{table}[h]
    \centering
    \caption{Comparison table for binary nucleation of butane mixed with \ce{CO2} at 51 K. The experimental data is taken from Ref~\cite{SI:Choudhury2026}.}
    \label{SItab:butane_binary}
    \adjustbox{max width=\linewidth}{%
\begin{tabular}{|c|c|c|c|c|c|c|c|c|} \hline\hline
         $C_{CO_2}/C_{Bu}$&  $J_{exp}$&  $J_{all}$&  $J^{\text{err}}_{all}$&  $J_{\text{Chp}}$&  $J^{\text{err}}_{\text{Chp}}$&  $J_{un}$&  $J^{\text{err}}_{un}$& $J_{HS}$\\
         Ratio&  $(10^{21}$&  $(10^{21}$&  $(10^{21}$&  $(10^{21}$&  $(10^{21}$&  $(10^{21}$&  $(10^{21}$& $(10^{24}$
\\
         &  m$^{-3}$s$^{-1})$&  m$^{-3}$s$^{-1})$&  m$^{-3}$s$^{-1})$&  m$^{-3}$s$^{-1})$&  m$^{-3}$s$^{-1})$&  m$^{-3}$s$^{-1})$&  m$^{-3}$s$^{-1})$& m$^{-3}$s$^{-1})$
\\\hline
         5.0&  4.0&  19.5&  5.0&  14.1&  3.6&  5.4&  1.4& 6.0\\
         10.0&  2.7&  34.9&  9.0&  29.2&  7.5&  5.7&  1.5& 6.2\\
         10.0&  2.3&  34.1&  8.8&  28.5&  7.3&  5.6&  1.4& 6.1\\
         13.2&  5.4&  43.4&  11.2&  37.8&  9.7&  5.6&  1.4& 6.1\\
         20.1&  3.5&  60.1&  15.5&  54.7&  14.1&  5.4&  1.4& 6.0\\
         25.0&  3.2&  74.3&  19.1&  68.8&  17.7&  5.5&  1.4& 6.1\\
         25.0&  4.4&  72.7&  18.7&  67.3&  17.3&  5.4&  1.4& 6.0\\
         30.0&  5.3&  87.4&  22.5&  81.9&  21.0&  5.6&  1.4& 6.1\\
         35.0&  10.5&  99.9&  25.7&  94.4&  24.3&  5.6&  1.4& 6.1\\
         35.0&  4.4&  97.1&  25.0&  91.7&  23.6&  5.4&  1.4& 6.0\\
         40.1&  7.3&  110.5&  28.4&  105.0&  27.0&  5.4&  1.4& 6.0\\
         45.0&  8.9&  118.4&  30.4&  113.1&  29.1&  5.3&  1.4& 5.8\\
         50.3&  6.7&  129.3&  33.2&  124.0&  31.9&  5.3&  1.4& 5.8\\ \hline\hline
    \end{tabular}}
\end{table}

\begin{table}[h]
    \centering
    \caption{Comparison table for binary nucleation of toluene mixed with \ce{CO2} at 55 K. The experimental data is taken from Ref~\cite{SI:Li2021}.}
    \label{SItab:tol_binary}
    \adjustbox{max width=\linewidth}{%
\begin{tabular}{|c|c|c|c|c|c|c|c|c|} \hline\hline
         $C_{CO_2}/C_{Tol}$&  $J_{exp}$&  $J_{all}$&  $J^{\text{err}}_{all}$&  $J_{Chp}$&  $J^{\text{err}}_{Chp}$&  $J_{un}$&  $J^{\text{err}}_{un}$& $J_{HS}$\\
         Ratio&  $(10^{22}$&  $(10^{21}$&  $(10^{21}$&  $(10^{21}$&  $(10^{21}$&  $(10^{20}$&  $(10^{20}$& $(10^{23}$
\\
         &  m$^{-3}$s$^{-1})$&  m$^{-3}$s$^{-1})$&  m$^{-3}$s$^{-1})$&  m$^{-3}$s$^{-1})$&  m$^{-3}$s$^{-1})$&  m$^{-3}$s$^{-1})$&  m$^{-3}$s$^{-1})$& m$^{-3}$s$^{-1})$
\\\hline
         0.0&  1.4&  2.0&  0.2&  0.0&  0.0&  20.0&  1.7& 17.1\\
         8.8&  0.9&  2.7&  0.2&  2.1&  0.2&  5.7&  0.5& 4.9\\
         17.5&  1.9&  4.8&  0.4&  4.3&  0.4&  5.7&  0.5& 4.9\\
         25.9&  3.7&  6.9&  0.6&  6.3&  0.5&  5.7&  0.5& 4.9\\
         32.9&  1.3&  2.4&  0.2&  2.3&  0.2&  1.6&  0.1& 1.4\\
         48.8&  1.7&  3.5&  0.3&  3.3&  0.3&  1.6&  0.1& 1.4\\ \hline\hline
    \end{tabular}}
\end{table}

\begin{table}[h]
    \centering
    \caption{Comparison table for binary nucleation of water mixed with \ce{CO2} at 57 K. The experimental data is taken from Ref~\cite{SI:Feusi2024}.}
    \label{SItab:water_binary}
    \adjustbox{max width=\linewidth}{%
\begin{tabular}{|c|c|c|c|c|c|c|c|c|} \hline\hline
         $C_{CO_2}/C_{H_2O}$&  $J_{exp}$&  $J_{all}$&  $J^{\text{err}}_{all}$&  $J_{Chp}$&  $J^{\text{err}}_{Chp}$&  $J_{un}$&  $J^{\text{err}}_{un}$& $J_{HS}$\\
         Ratio&  $(10^{21}$&  $(10^{21}$&  $(10^{21}$&  $(10^{21}$&  $(10^{21}$&  $(10^{20}$&  $(10^{20}$& $(10^{24}$
\\
         &  m$^{-3}$s$^{-1})$&  m$^{-3}$s$^{-1})$&  m$^{-3}$s$^{-1})$&  m$^{-3}$s$^{-1})$&  m$^{-3}$s$^{-1})$&  m$^{-3}$s$^{-1})$&  m$^{-3}$s$^{-1})$& m$^{-3}$s$^{-1})$
\\\hline
         9.5&  3.5&  6.2&  1.0&  5.7&  1.0&  4.4&  0.7& 2.6\\
         23.8&  6.2&  8.6&  1.5&  8.3&  1.4&  2.6&  0.4& 1.5\\
         35.6&  3.7&  5.6&  0.9&  5.5&  0.9&  1.1&  0.2& 0.7\\
         39.6&  13.2&  14.0&  2.4&  13.7&  2.3&  2.6&  0.4& 1.5\\
         39.6&  8.8&  13.8&  2.3&  13.6&  2.3&  2.6&  0.4& 1.5\\
         55.4&  20.5&  18.4&  3.1&  18.2&  3.1&  2.6&  0.4& 1.5\\
         59.4&  8.4&  9.4&  1.6&  9.3&  1.6&  1.2&  0.2& 0.7\\
         83.1&  14.2&  12.5&  2.1&  12.4&  2.1&  1.2&  0.2& 0.7\\ \hline\hline
    \end{tabular}}
\end{table}

\clearpage
\subsection{Beyond dimers}
\begin{table}[h]
    \centering
    \caption{Comparison between experimental ($J_{exp}$) and theoretically predicted nucleation rates using dimers ($J_{\text{dimers}}$) and trimers ($J_{\text{trimers}}$) as critical clusters from collision limited to barrier controlled nucleation of \ce{CO2}. The experimental data is obtained from Ref~\cite{SI:Dingilian2021}.}
    \label{SItab:beyond_dymer}
    \adjustbox{max width=\linewidth}{%
\begin{tabular}{|c|c|c|c|c|c|} \hline \hline
         $T$&  $J_{exp}$&  $J_{\text{trimers}}$&  $J^{\text{error}}_{\text{trimers}}$&  $J_{\text{dimers}}$& $J^{\text{error}}_{\text{dimers}}$\\
 (K)& (cm$^{-3}$s$^{-1}$)& (cm$^{-3}$s$^{-1}$)& (cm$^{-3}$s$^{-1}$)& (cm$^{-3}$s$^{-1}$)&(cm$^{-3}$s$^{-1}$)\\ \hline
         31.2&  2.5E+15&  1.7E+15&  3.8E+14&  1.8E+15& 3.7E+14\\
         44.4&  1.7E+16&  3.5E+16&  1.1E+16&  4.4E+16& 9.3E+15\\
         49.3&  4.7E+16&  9.0E+16&  3.4E+16&  1.4E+17& 3.0E+16\\
         56.4&  1.4E+17&  3.7E+17&  1.6E+17&  8.7E+17& 1.9E+17\\
         62.9&  5.5E+17&  3.9E+18&  1.6E+18&  1.6E+19& 3.5E+18\\ \hline \hline
    \end{tabular}}
\end{table}

\clearpage

\end{document}